\documentclass[12pt,aps,preprint,pra]{revtex4}
\usepackage{amsmath}
\usepackage{amssymb}
\usepackage{graphicx}

\begin{document}

\title{Weak interaction between germanene and GaAs(0001) by H intercalation: A route to exfoliation}

\author{T. P. Kaloni and U. Schwingenschl\"ogl}

\email{udo.schwingenschlogl@kaust.edu.sa,+966(0)544700080}

\affiliation{Physical Science \& Engineering Division, KAUST, Thuwal 23955-6900, Kingdom of Saudi Arabia}

\begin{abstract}
Epitaxial germanene on a semiconducting GaAs(0001) substrate is studied by ab initio calculations. The germanene-substrate interaction is found to be strong for direct contact but can be substantially reduced by H intercalation at the interface. Our results indicate that it is energetically possible to take the germanene off the GaAs(0001) substrate. While mounted on the substrate, the electronic structure shows a distinct Dirac cone shift above the Fermi energy with a splitting of 175 meV. On the other hand, we find for a free standing sheet a band gap of 24 meV, which is due to the intrinsic spin orbit coupling. 
\end{abstract}

\maketitle

\section{Introduction}
Graphene is two-dimensional (2D) honeycomb network of carbon atoms, it is currently a material 
of interest for many researcher due to its unique electronic properties, which is proposed to be a 
potential candidate for future nanoelectronic devices application \cite{geim}. 
However, the mass production as well as band gap opening in graphene is a really a 
challenge as a result researcher searching new materials which can be 
a counterpart of graphene. Recent years, the electronic properties of 2D hexagonal 
lattices of silicon and Germanium also named as silicene and germanene, 
respectively, have been proposed and studied as a potential alternative of graphene 
\cite{yang,Eriksson,Topsakal}. Growth of silicene and its derivatives 
experimentally has been demonstrated on Ag and ZrB$_2$ substrates \cite{padova,vogt,ozaki}. 
However, the accurate measurements of the materials properties 
are difficult on a metallic substrate and still its questionable whether it could growth on a 
less interacting substrates, like semiconductors. Carbon (C), silicon (Si), and Germanium (Ge) 
belongs to the same group in the periodic table. Whereas, Si and Ge have a larger ionic radius, 
which promotes $sp^3$ hybridization. The possible mixture of $sp^2$ and $sp^3$ hybridization in 
silicene and germanene results in a prominent buckling of 0.46 \AA\ and 0.68 \AA, 
wich opens an electrically tunable band gap \cite{falko,Ni}. Hence, its a clearly huge 
advantage as compared to graphene. 

On the other hand unlike silicene, germanene is proposed to be a gapless semiconductor 
\cite{yang,Eriksson,Topsakal}. The first principles calculations 
of the electronic properties of germanene was studied and showed that germanene behaves like a 
metal \cite{Eriksson,Houssa}. It was also noted that in-plane biaxial compressive strain turns 
germanene into a gapless semiconductor, by remain intact the linear energy dispersions at the 
K and K$'$ points. The Fermi velocity is found to be independent on the strain and the magnitude 
is 1.7$\cdot$10$^6$ m/s, which is comparable to that of graphene ($1\cdot10^6$) \cite {Avouris} and 
silicene ($1.3\cdot10^6$ \cite{vogt}). The intrinsic spin orbit coupling in Ge is stronger than that 
for C and Si atoms. The magnitude of the spin orbit coupling for Ge, Si, and C is 46.3 meV, 4 meV, 
and $1.3\cdot10^{-3}$ \cite{Liu1}, respectively. It was demonstrated that germanene would be able 
to realize the quantum spin hall effect with a sizable band gap at the Dirac points due to 
stronger spin orbit coupling and the higher buckling as compared to silicene \cite{Liu,Liu1}. 
Hence, germanene is a promising candidate to construct potential spintronic devices. More recently, 
by adopting a first principles method the electronic properties of germanene with chemisorption of 
F, Cl, Br and I has been studied \cite{ma}. More importantly, the intrinsic SOC band gap in germanene 
is enhanced by absorption of Cl, Br, and I with the order of 86 meV, 237 meV and 162 meV, 
which indeed higher than that for pristine germanene.

Moreover, it has been reported that the bulk Ge in the form of graphite-like Ge, the structural 
energy continues to increase with increasing the interlayer distance between two neighboring 
germanene layers \cite{wang}. 
Such a behavior is different from that of graphene. As a result it would be 
difficult to form a layered structure in germanene like in graphite. At this point its a open question: 
whether germenene can be grown on any kinds of appropriate substrate? Note that choice of the proper 
substrate is the really important. However, the epitaxial growth of the Ge films on GaAs(100) substrates has been 
demonstrated experimentally and it has been proposed that such a thin film is potential for 
Ge-Source/Drain GaAs MOSFETs \cite{luo}. On the other hand, the growth of GaAs on Ge has been 
verified experimentally \cite{Galiana}. Hence, the growth of germanene is the most important issue 
with the proper substrate, to achieve this goal, we therefore, in this letter, study the growth of 
germanene on the GaAs(0001) substrate and argue that germanene can be easily grown 
and take off from GaAs(0001) substrate. Based on the experimental data of the Refs. \cite{luo} 
and \cite{Galiana}, it is a strong beleive that the GaAs(0001) 
would be the appropriate substrate for epitaxial growth of germanene. Further, 
we study the electronic properties of germanene with and without substrate. 
We also calculate the electronic band structure of the germanene by the incorporation of spin orbit coupling.

\section{Computational details}
We have carried out density functional theory in the generalized gradient approximation using PWSCF code \cite{paolo}. 
The van der Waals interaction \cite{grime,jmc} is taken into account for dispersion correction. Plane wave cutoff energy 
of 816 eV is used. A Monkhorst-Pack $8\times8\times1$ k-mesh is employed for optimizing the 
geometry and a refined $16\times16\times1$ k-mesh is used increase the accuracy of the self-consistency calculation. 
The in-place lattice parameters of the supercell are set to the bulk GaAs value of $2\cdot3.91$ \AA\ with a vaccum of 
16 \AA. Atomic positions are relaxed until an energy convergence of 10$^{-6}$ eV and a force 
convergence of $3\cdot10^{-4}$ eV/\AA\ are reached. We deal with a supercell consisting of $2\times2$ germanene (8 Ge atoms)
on the top of a Ga-terminated GaAs(0001) slab, which contains 3 layers (12 Ga atoms) of Ga and 3 layers of As (12 As atoms). 
This supercell fits on a $2\times2$ supercell of GaAs with a lattice mismatch of 3.6\%. The dangling electron on the top and bottom 
layers of the substrate is saturated by H atoms.

\section{Structural analysis}
\begin{table*}[h]
\begin{tabular}{|c|c|c|c|c|c|c|}
\hline
Configuration &Ge$-$Ge (\AA) &Ga$-$H (\AA)       &$\theta$   &Buckling (\AA)       &$E_b$ (meV) \\                                                         
\hline                                                     
\hline
(a)           &2.43          &$-$                &109-110    &0.75-0.80              &  577 \\
\hline
(b)           &2.41          &1.57               &111        &0.71-0.72              &  86 \\
\hline
\end{tabular}
\caption{Selected bond lengths, Ge bond angle (in degrees), buckling, and binding energy.}
\end{table*}

Firstly, a monolayer of germanene on Ga-faced GaAs(0001) substrate is studied, see Fig.\ 1. 
We also studied the growth on As-faced GaAs(0001) substrate but find a huge structural distortions 
which we think the experimentally it would be difficult to achieve. Therefore, in the following 
section, we focus on Ga-faced GaAs(0001). In the configuration addressed in Fig.\ 1(a), the buckling 
of the germanene monolayer is found to be larger (0.75-0.80 \AA) than that of the pristine germanene 
($\sim$0.68 \AA) \cite{Ni}. The larger buckling is realized due to the interaction with the substrate. 
we obtain a 2\% elongation of the Ge$-$Ge bond length due to the buckling has increased, see Table I. 
The bond angle between neighboring Si atoms ( $\theta$) amounts to 109$^{\circ}$-110$^{\circ}$, which 
reflects that the germanene support rather the $sp^3$ hybridizations. The obtained $\theta$ is lower 
than to the pristine germanene of 112.3$^{\circ}$ \cite{ma}.

\begin{figure*}[h]
\includegraphics[width=0.8\textwidth]{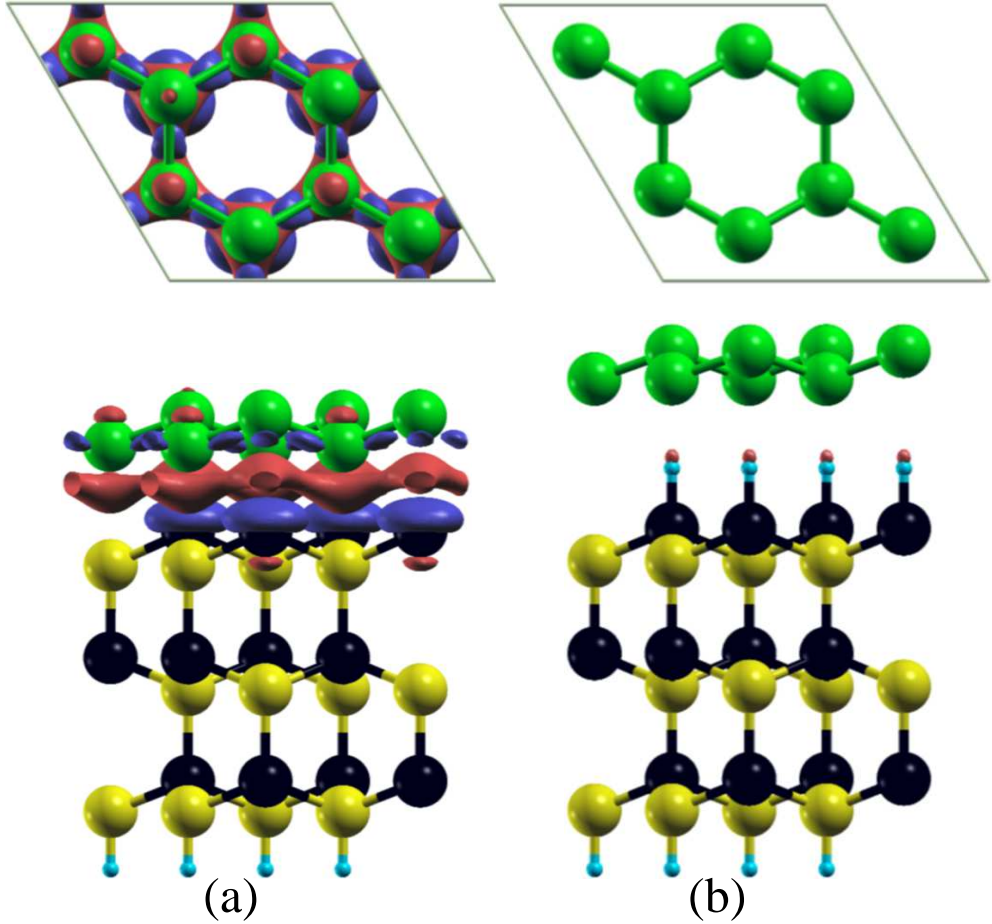}
\caption{Relaxed structures of the configurations under consideration with charge density difference isosurfaces 
(a) without and (b) with H passivated Ga layer. The isosurfaces shown correspond to the small isovalues of 
$\pm3\cdot10^{-3}$ electrons/$a^3_B$. The green, black, yellow, and blue spheres represent Ge, Ga, As, and H, respectively.}
\end{figure*}

In the configuration shown in Fig.\ 1(b). The dangling Ga electrons are saturated by H atoms. The H 
atoms forms covalent bonds with Ga atoms with a bond length of 1.57 \AA. Indeed, the process of 
H-intercalation in graphene on SiC(0001) was carried out in a chemical vapor deposition reactor 
on atmosphere of palladium-purified ultrapure molecular hydrogen \cite{riedl}. This technique is 
similar to hydrogen etching and hydrogen passivation \cite{Tsuchida} of SiC surfaces. The developed 
morphology after H intercalation has been observed by the Angle-resolved photoelectron spectroscopy, 
low energy electron diffraction and microscopy low energy electron diffraction. By relying on these 
experimental data, we believe that the hydrogenation of GaAs(0001) can be achieve easily. On our 
calculations, the observed magnitude of the buckling and the angle $\theta$ are 0.71-0.72 \AA\ 
and 111$^{\circ}$, very close to the pristine germanene. The Ge$-$Ge bond found to be about 1\% 
longer. All these parameters are indicate of the weak interaction of the substrate with germanene 
layer as compared to the previous case. In the configuration in Fig.\ 1(a) the charge density isosurfaces, 
its is clear that there is a huge amount of charge redistribution between the interface of Ge layer and 
the top GaAs(0001) layer. This indicates that Ge layer is hugely affected by the substrate, next section 
we will prove this by the calculated binding energy. The fact that there is essentially almost no charge 
redistribution in between Ge layer and the substrate due to the H intercalation as a result Ge layer is 
weakly influenced by the substrate in the configuration addressed in Fig.\ 1(b). Other way around, in 
graphene by the effect of H intercalation the buffer layer decoupled from the substrate and form a 
monolayer graphene also called eptaxially growth of graphene on SiC(0001) substrate \cite{riedl}. 

\section{Electronic structure}
\begin{figure*}[h]
\includegraphics[width=1\textwidth,clip]{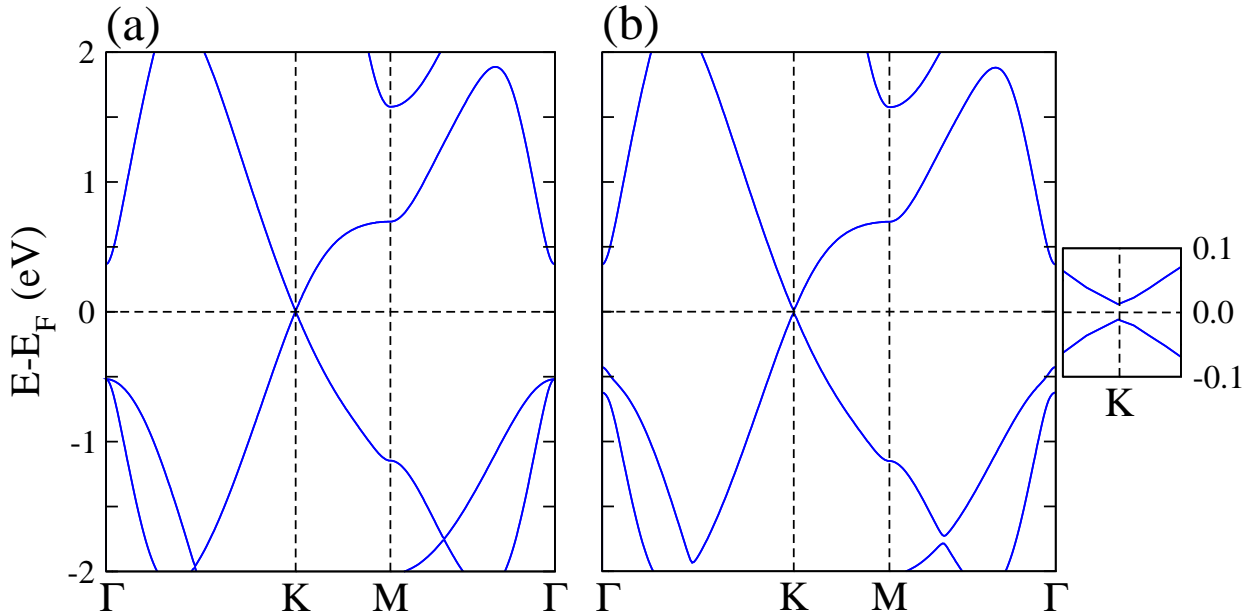}
\caption{Electronic band structures of free standing germanene (a) without and (b) with SOC.}
\end{figure*}
In order to get insight whether germanene can be take off from the GaAs(0010) substrate we 
calculate the binding energy $E_b$ between the germanene and the substrate. 
We use the mathematical expression $E_b=E_{T}-E_{GaAs}-E_{Ge}$, $E_{T}$ is the 
total energy of the whole system, $E_{GaAs}$ the total energy of the GaAs(001) 
substrate, and $E_{Ge}$ is the total energy of the germanene layer. We obtain values 
$577$ meV and $86$ meV for the configuration in Fig.\ 1(a) and 1(b), respectively.
Note that these values refer to per Ge atoms. The monolayer germanene without H 
passivated dangling bonds of Ga thus still substantially bound to the substrate,
as a result it would be difficult to take off to make it free standing. By the H 
intercalation it strongly reduces the interaction between germanene and the substrate. 
With an average binding energy of only 86 meV per Ge atom we can expect that in this 
configuration the germanene can be separated from the substrate to experimentally obtain 
free standing germanene. The experimentally observed data for the binding energy of the 
graphitic layers is ranges 47-57 meV \cite{Zacharia}, this value could be higher when the 
graphene grown on SiC(0001) and other meta substrate due to stronger interaction with the 
substrate, while in graphite, graphene layers are rather weakly bound. Recently, silicene 
has been grown on Ag(111) substrate \cite{Enriquez} and claims that the silicene sheet is 
weakly bound to Ag substrate, with average binding energy per silicon atom is 460 meV. It 
means silicene could be take off from Ag(111) substrate to make it free standing. Based on 
these data, we believe that germanene could be easy take off from the GaAs(0001) substrate after growth.

\begin{figure}[h]
\includegraphics[width=0.8\textwidth,clip]{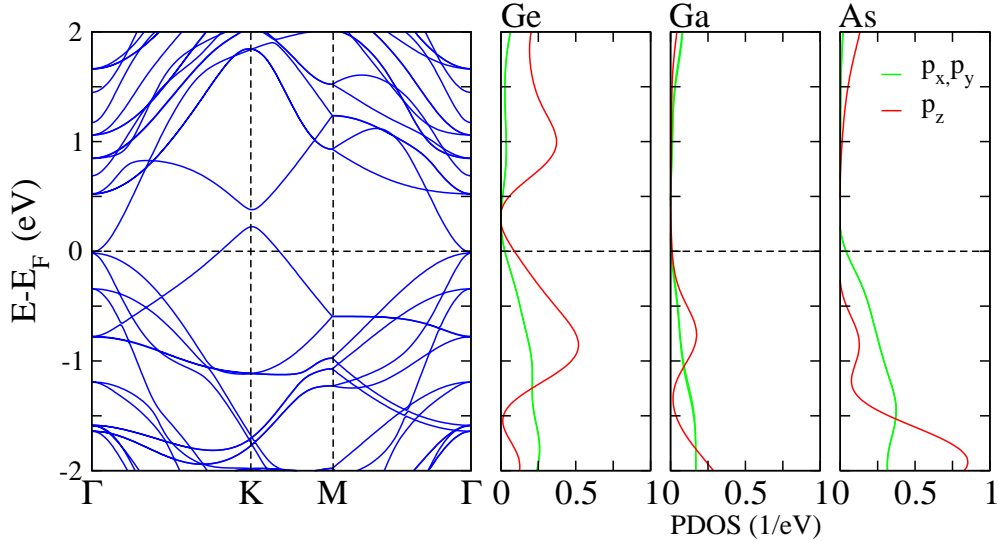}
\caption{Electronic structures of germanene on GaAs(0001) substrate with corresponding density of states for 1 Ge atom atom at the top of the H atom, 1 Ga atom bound with H atom on the top layer of the substrate, and 1 As atom bound with the Ga atom in the same plane.}

\end{figure}
The electronic band structure of the free standing germanene is addressed in Fig.\ 2. 
We find that germanene behaves like a metal with the the linear energy dispersion at the K point, 
see Fig.\ 2(a), consistent with the previous reports \cite{yang,Eriksson,Topsakal}. Like graphene 
and silicene the $\pi$ and $\pi^*$ bands of the Dirac cone are contributed by the $p_z$ orbitals of Ge. 
However, we find a band gap of 24 meV (see intert in Fig.\ 2(b)) by taken into account an intrinsic 
spin orbit coupling on our calculations, consistent with the available reports. This indicates 
that the spin orbit coupling can not neglected in germanene. This material could have a great application 
in electronic devices. By applying an external electric field, the magnitude of the band gap could be 
enhanced easily because the external electric field break the sublatice symmetry as a result the band 
gap due to spin orbit coupling could be increases. Such a effect has been observed in case of silicene 
\cite{falko,Ni}. The germanene is weakly interact to the GaAS(0001) as a result the Dirac cone shift 
above the Fermi level by about 250 meV and splitted with the gap of 150 meV, see Fig.\ 3. The projected 
densities of states shown in right side of the Figure indicates that the $\pi$ and $\pi^*$ bands of the 
Dirac cone is contributed from the $p_z$ orbitals of the Ge atoms with a minor contribution from the 
$p_x$ and $p_y$ orbitals as expected and the contribution from the substrate atoms are almost not seen 
at the Dirac cone. This confirms that the splitted Dirac cone is purely comes from the germanene.  

\section{Conclusion}
In summary, we have discussed the structural analysis, electronic properties and the chemical bonding of 
germanene on GaAs(0001) substrate. We find a substantial bonding to the substrate for monolayer germanene, 
while the H atom intercalation strongly reduces the interaction with the substrate. Our results point out that in 
the latter case the germanene can be removed easily from the substrate, which opens the way to take off the 
germanene and make it to free standing. Furthermore, we have studied the electronic structure of the germanene 
on the substrate and find that the Dirac cone splitted by 150 meV and shift about 250 meV 
above the Fermi level as compared to pristine germanene. The electronic structure of the free standing 
germanene shows that the Dirac cone lies in the Fermi level at the K point. However, the intrinsic spin 
orbit coupling opens a band gap of the 24 meV, which is huge as compared to graphene and silicene. This 
indicative of the spin orbit coupling can not be neglected in germanene unlike graphene. According to or 
study, growth of germanene on a semiconducting substrate is possible and highly useful for device applications.

\end{document}